\documentclass[aps,pra,floatfix,superscriptaddress]{revtex4}
\usepackage{graphicx}
\usepackage{epstopdf}

\usepackage{dcolumn}
\usepackage{amsmath}
\usepackage{amssymb}
\usepackage{dsfont}
\usepackage{bm}

\begin{document}

\title{STABILITY OF POINT SPECTRUM FOR THREE-STATE QUANTUM WALKS ON A LINE}

\author{M. \v Stefa\v n\'ak\email[correspondence to:]{martin.stefanak@fjfi.cvut.cz}}
\affiliation{Department of Physics, Faculty of Nuclear Sciences and Physical Engineering, Czech Technical University in Prague, B\v
rehov\'a 7, 115 19 Praha 1 - Star\'e M\v{e}sto, Czech Republic}

\author{I. Bezd\v ekov\'a}
\affiliation{Department of Physics, Faculty of Nuclear Sciences and Physical Engineering, Czech Technical University in Prague, B\v
rehov\'a 7, 115 19 Praha 1 - Star\'e M\v{e}sto, Czech Republic}

\author{I. Jex}
\affiliation{Department of Physics, Faculty of Nuclear Sciences and Physical Engineering, Czech Technical University in Prague, B\v
rehov\'a 7, 115 19 Praha 1 - Star\'e M\v{e}sto, Czech Republic}

\author{S. M. Barnett}
\affiliation{School of Physics \& Astronomy, University of Glasgow, Glasgow G12 8QQ, Scotland, U.K.}

\begin{abstract}
Evolution operators of certain quantum walks possess,
apart from the continuous part, also point spectrum. The existence
of eigenvalues and the corresponding stationary states lead to
partial trapping of the walker in the vicinity of the origin. We
analyze the stability of this feature for three-state quantum walks
on a line subject to homogenous coin deformations. We find two
classes of coin operators that preserve the point spectrum. These
new classes of coins are generalizations  of coins found previously
by different methods and shed light on the rich spectrum of coins
that can drive discrete-time quantum walks.
\end{abstract}

\date{\today}

\maketitle


\section{Introduction}

Quantum walks \cite{aharonov} have become quite popular in the last few years. This is motivated by their potential applications in quantum information
theory \cite{santha}, statistical physics \cite{werner:random:coin,joye} and transport theory \cite{qw:transport}. Additional interest in quantum walks was stimulated by the now
considerable number of experiments \cite{karski,schmitz,Schreiber,Zahringer,Broome,OBrien,and:2dwalk:science,sansoni} which have demonstrated the basic
properties of quantum walks. They have shown in an impressive way
the quantum coherence which is needed for their realization. Among
the basic effects associated with quantum walks is the fast
spreading of the walker across the underlying grid.

The key role in the analysis of the quantum walk plays the determination of the spectrum of the unitary evolution operator. For quantum walks with homogeneous coin on infinite lattice
one can employ Fourier analysis \cite{ambainis}, which reduces this problem to that of finding the eigenvalues of a finite-size matrix dependent on the wave-number $k$. The ballistic spreading of the quantum walk can be deduced from the
analogy with wave theory. The continuous spectrum of the evolution
operator corresponds to the $k$-dependent eigenvalues which can be described by dispersion relations. This allows one to find the group velocity and its distribution \cite{Grimmett} which determines the propagation of the wave packets. The peaks in the probability
distribution of the quantum walk propagate at constant rate given by
the maximum of the group velocity \cite{kempf}. However, the evolution operators
of certain quantum walks also have a non-empty point spectrum, which is represented by $k$-independent eigenvalues. In such a case, the evolution does not consist of purely ballistic spreading. Indeed, as the walker spreads through the lattice its
wave-function overlaps with the stationary states. The walker is
therefore partially trapped in the vicinity of the origin. This feature, also known as localization, was found in the three-state walk on a line with the Grover coin
operator \cite{konno:loc:2005,konno:loc:2005b}, where the evolution
operator has one eigenvalue equal to unity. Similarly, Grover walk
on a square lattice also has a point spectrum \cite{konno:loc:2004}
consisting of $\pm 1$. This can be exploited for a number of effects. The form of the spectrum can be used to sculpture the shape of the walker's wave packet, the walker can be trapped at particular position and
can also lead to the effect of full revival \cite{stef:rev}, where the walker's wave-packet undergoes a periodic time-evolution. It may be anticipated, however, that the presence of the point spectrum will be highly sensitive to the choice of the coin
operator. Even a small perturbation in a wrong direction can eliminate
the eigenvalues. This can be crucial for experimental realizations
of such quantum walks, where the imperfections in all operations has
to be taken into account. In \cite{konno:loc:2008} the authors have
analyzed a one-parameter modification of the Grover walk on a square
lattice which preserves the point spectrum. The coin parameter
controls the rate at which the particle spread through the lattice.
We have extended this idea to three-state walk \cite{iva:cont:def}
on a line and found two one-parameter families of walks with point
spectrum. Their coin operators are constructed as either eigenvalue
or eigenvector deformations of the Grover coin. It is not
clear, however, whether the two sets exhaust all possible three-state walks with
point spectrum. The present paper aims to address this issue. The
determination of coin families with a point spectrum contribute
significantly to the classification of coins with respect to their
physical properties, i.e. to localizing and non-localizing coins.
Even though this is a very crude classification it certainly helps and
in addition it simplifies experimental considerations when the wave
packet propagating as a quantum walk is of interest.

The paper is organized as follows: In Section~\ref{sec2} we find the
conditions on the coin operator which guarantees that the evolution
operator of the quantum walk has a point spectrum. We solve these
requirements in Section~\ref{sec3} with the help of a particular
parametrization of the unitary group. We find three trivial
solutions and two non-trivial ones. In Section~\ref{sec4} we analyze the dependence of
the rate of spreading of the walk through the lattice on the remaining coin parameters. Finally, we study the trapping of
the walker in Section~\ref{sec5}. We conclude and present an outlook
in Section~\ref{sec6}.

\section{Characteristic equation and conditions on the coin operator}
\label{sec2}

We consider a three-state discrete-time quantum walk on a line with a homogeneous coin operator $\hat{C}$. We denote the basis coin states as $|L\rangle$, $|S\rangle$ and $|R\rangle$, which correspond to the step to the left, staying at the present position and the step to the right. The simplest way to solve the dynamics is to analyze it in the momentum representation \cite{ambainis}. In the Fourier representation the evolution operator has the form
\begin{equation}
\label{evol:op}
\tilde{U}(k) = D\left(e^{-i k},1,e^{i k}\right)\cdot C,
\end{equation}
where $D$ denotes a diagonal matrix and $C$ is the matrix representation of the coin operator with matrix elements
\begin{equation}
C_{ij} = \langle i|\hat{C}|j\rangle, {\rm\ with\ } i,j = L,\ S,\ R. \end{equation}
We are interested in quantum walks which show the localization effect. This feature corresponds to the fact that the evolution operator in the Fourier representation (\ref{evol:op}) has an eigenvalue independent of $k$. Note that if (\ref{evol:op}) has two eigenvalues independent of $k$, the third one has to be also constant. This follows immediately from the fact that the determinant of (\ref{evol:op}) is the same as determinant of $C$ which is independent of $k$. The case when the evolution operator (\ref{evol:op}) has all three eigenvalues independent of $k$ leads to a trivial quantum walk with no spreading. Let us therefore assume that only one eigenvalue of the evolution operator is independent of $k$. Then we can always put the eigenvalues into the form
\begin{equation}
\lambda_0 = e^{i\varphi},\quad \lambda_{1,2}(k) = e^{\pm i \omega(k)},
\end{equation}
simply by multiplying the coin operator by a global phase factor, which does not influence the overall dynamics. The function $\omega(k)$ has to be real for all $k$, since the evolution operator $\tilde{U}(k)$ is unitary and its eigenvalues must be of modulus 1. Consider the characteristic equation
\begin{equation}
\rm{det}\left(\tilde{U}(k) - \lambda \right) = (\lambda_0-\lambda)(\lambda_1-\lambda)(\lambda_2-\lambda) = 0.
\end{equation}
The terms with same power of $\lambda$ on the left and the right
hand side of the equation give the following relations
\begin{itemize}
\item[$\lambda^0$:] $\qquad e^{i\varphi} = {\rm det}\ C$
\item[$\lambda^1$:] $\qquad 1 + e^{i\varphi}\left(e^{i\omega(k)} + e^{-i\omega(k)}\right) = m_L e^{ik} + m_S + m_R e^{-ik}. $
\item[$\lambda^2$:] $\qquad e^{i\varphi} + e^{i\omega(k)} + e^{-i\omega(k)} = C_{LL}e^{-ik} + C_{SS} + C_{RR}e^{ik}$
\end{itemize}
Here we have denoted by $m_i$ the minors of the coin operator, i.e.
\begin{equation}
m_L = \rm{det}\left(
                \begin{array}{cc}
                  C_{SS} & C_{SR} \\
                  C_{RS} & C_{RR} \\
                \end{array}
              \right),\quad m_S = \rm{det}\left(
                \begin{array}{cc}
                  C_{LL} & C_{LR} \\
                  C_{RL} & C_{RR} \\
                \end{array}
              \right),\quad m_R = \rm{det}\left(
                \begin{array}{cc}
                  C_{LL} & C_{LS} \\
                  C_{SL} & C_{SS} \\
                \end{array}
              \right).
\end{equation}
The third equation leads to the dispersion relations determining the $\omega(k)$ in the form
\begin{equation}
\label{disp1}
2\cos\omega(k) = C_{LL}e^{-ik} + C_{SS} + C_{RR}e^{ik} - e^{i\varphi}.
\end{equation}
This function has to be real for all $k$, which is only possible if
\begin{equation}
C_{LL} = C_{RR}^* = \rho e^{i\gamma}, \quad C_{SS} = e^{i\varphi} - 2 \mu,\quad \rho,\ \gamma,\ \mu\in\mathds{R},
\end{equation}
where the star denotes the complex conjugation. The dispersion
relations then attain a simple form
\begin{equation}
\label{omega}
\cos\omega(k) = \rho\cos(k-\gamma) - \mu.
\end{equation}
Note that they are fully determined by the diagonal elements of the coin operator. Moreover, if $\rho=0$, i.e. when $C_{LL}=C_{RR} = 0$, then $\omega$ is constant. In such a case the evolution operator has purely point spectrum and the quantum walk is trivial - it does not spread at all.

Let us now consider the terms with $\lambda^1$ which lead us to the dispersion relations in the form
\begin{equation}
2\cos\omega(k) = e^{-i\varphi}\left(m_L e^{ik} + m_S + m_R e^{-ik} -1\right).
\end{equation}
Comparing this formula with (\ref{disp1}) we find the following conditions involving the off-diagonal elements of $C$
\begin{eqnarray}
\label{a11} C_{LL} & = & e^{-i\varphi}\left(C_{LL}C_{SS} - C_{LS}C_{SL}\right),\\
\label{a33} C_{RR} & = & e^{-i\varphi}\left(C_{RR}C_{SS} - C_{SR}C_{RS}\right) = C_{LL}^*,\\
\label{a22} C_{SS} - e^{i\varphi} & = & e^{-i\varphi}\left(C_{LL}C_{RR} - C_{LR}C_{RL} - 1\right).
\end{eqnarray}
Moreover, the matrix $C$ has to be unitary.

\section{Parametrization of the unitary group}
\label{sec3}

In order to find coins which satisfy the conditions (\ref{a11})-(\ref{a22}) we first parameterize the three-dimensional unitary group, which has a dimension nine, in the following way \cite{jarlskog}
\begin{equation}
C = D\left(e^{i \alpha_1},e^{i \alpha_2},e^{i \alpha_3}\right)\cdot V \cdot D\left(e^{i \beta_1},e^{i \beta_2},e^{i \beta_3}\right).
\end{equation}
Here the matrix $V$ is the quark mixing matrix
\begin{equation}
V = \left(
\begin{array}{ccc}
  c_{12} c_{13} &  c_{13} s_{12} & e^{-i \delta} s_{13} \\
 -c_{23} s_{12}-e^{i \delta} c_{12} s_{13} s_{23} &  c_{12} c_{23}-e^{i \delta} s_{12} s_{13} s_{23} &  c_{13} s_{23} \\
 s_{12} s_{23}-e^{i \delta} c_{12} c_{23} s_{13} & - c_{12} s_{23} - e^{i \delta} c_{23} s_{12} s_{13} & c_{13} c_{23}
\end{array}
\right),
\end{equation}
familiar from the Standard model \cite{nakamura}. For brevity we have used the
notation
\begin{equation}
c_{ij} = \cos\theta_{ij}, s_{ij} = \sin\theta_{ij}.
\end{equation}
The mixing matrix has four real parameters $\theta_{12},\theta_{13},\theta_{23}$ and $\delta$. The five remaining independent parameters are
\begin{equation}
\gamma_1 = \alpha_1+\beta_1,\ \gamma_2 = \alpha_1+\beta_2,\ \gamma_3 = \alpha_1+\beta_3,\ \gamma_4 = \alpha_2+\beta_1,\ \gamma_5 = \alpha_3+\beta_1.
\end{equation}
With this parametrization a general 3x3 unitary matrix is given by
\begin{equation}
C = \left(
\begin{array}{ccc}
 e^{i \gamma_1} c_{12} c_{13} & e^{i \gamma_2} c_{13} s_{12} & e^{-i (\delta-\gamma_3)} s_{13} \\
 -e^{i \gamma_4} \left(c_{23} s_{12}+e^{i \delta} c_{12} s_{13} s_{23}\right) & e^{-i (\gamma_1-\gamma_2-\gamma_4)} \left(c_{12} c_{23}-e^{i \delta} s_{12} s_{13} s_{23}\right) & e^{-i (\gamma_1-\gamma_3-\gamma_4)} c_{13} s_{23} \\
 e^{i \gamma_5} \left(s_{12} s_{23} - e^{i \delta} c_{12} c_{23} s_{13}\right) & -e^{-i (\gamma_1-\gamma_2-\gamma_5)} \left(c_{12} s_{23} + e^{i \delta} c_{23} s_{12} s_{13}\right) & e^{-i (\gamma_1-\gamma_3-\gamma_5)} c_{13} c_{23}
\end{array}
\right).
\end{equation}
Note that the determinant of $C$ equals
\begin{equation}
{\rm det}\ C = e^{i\varphi} = e^{-i(\gamma_1-\gamma_2-\gamma_3-\gamma_4-\gamma_5)}.
\end{equation}

Let us now turn to the requirements for the non-empty point spectrum of the evolution operator. The relations (\ref{a11}) and (\ref{a33}) lead to the condition
\begin{equation}
\label{cond}
c_{13} \left(c_{12} - e^{i (\gamma_3 + \gamma_5)} c_{23}\right) = 0.
\end{equation}
This is satisfied in the following cases:
\begin{enumerate}
\item $c_{13} = 0$, i.e. $\theta_{13} = \frac{\pi}{2}$  - trivial solution, no dynamics

The coin operator has the form
\begin{equation}
C = \left(
\begin{array}{ccc}
0 & 0 &  e^{-i (\delta-\gamma_3)} \\
 -e^{i \gamma_4} \left(c_{23} s_{12} + e^{i \delta} c_{12} s_{23}\right) & e^{-i (\gamma_1-\gamma_2-\gamma_4)} \left(c_{12} c_{23} - e^{i \delta} s_{12} s_{23}\right) & 0 \\
 e^{i \gamma_5} \left(s_{12} s_{23} - e^{i \delta} c_{12} c_{23}\right) & -e^{-i (\gamma_1-\gamma_2-\gamma_5)} \left(c_{12} s_{23} + e^{i \delta} c_{23} s_{12}\right) & 0
\end{array}
\right).
\end{equation}
Since $C_{LL}=C_{RR}= 0$, the evolution operator does not have a continuous spectrum.

We note that the alternative choice of $\theta_{13}=-\frac{\pi}{2}$ results in an equivalent matrix (in the sense of the properties of the quantum walk). The same will apply to other solutions of equation (\ref{cond}) given bellow. We will therefore always treat only one possible choice of the angles in the range $(-\pi,\pi)$.

\item $c_{12} = c_{23} = 0$, i.e. $\theta_{12} = \theta_{23} = \frac{\pi}{2}$ - trivial solution, no dynamics

The coin operator has the form
\begin{equation}
C = \left(
\begin{array}{ccc}
 0 & e^{i \gamma_2} c_{13} & e^{-i (\delta-\gamma_3)} s_{13} \\
 0 & - e^{i (\delta-\gamma_1+\gamma_2+\gamma_4)} s_{13} & e^{-i (\gamma_1-\gamma_3-\gamma_4)} c_{13} \\
 e^{i \gamma_5} & 0 & 0
\end{array}
\right).
\end{equation}
Since $C_{LL}=C_{RR}= 0$, the evolution operator does not have a continuous spectrum.

\item $\gamma_3 = -\gamma_5$, $c_{12} = c_{23}$

From the equation (\ref{a22}) follows the condition
\begin{equation}
(\sin(\gamma_1 - \gamma_2 - \gamma_4) - \sin(\delta - \gamma_1 + \gamma_2 + \gamma_4) s_{13}) s_{23} = 0.
\end{equation}
This requires that one of the following is satisfied:
\begin{enumerate}
\item $s_{23} = 0$, i.e. $\theta_{23} = 0$ - trivial solution, decoupling

\begin{equation}
C = \left(
\begin{array}{ccc}
 e^{i \gamma_1} c_{13} & 0 & e^{-i (\delta+\gamma_5)} s_{13} \\
 0 & e^{-i (\gamma_1-\gamma_2-\gamma_4)} & 0 \\
 -e^{i (\delta+ \gamma_5)} s_{13} & 0 & e^{-i \gamma_1} c_{13}
\end{array}
\right)
\end{equation}
In this case the state of the coin $|S\rangle$ is decoupled from the other two states $|L,R\rangle$. The walk reduces to a two-state walk and the $|S\rangle$ component of the initial state remains at the origin.

\item $\delta=0$, $\gamma_1 = \gamma_2+\gamma_4$ - nontrivial solution

In this case the coin operator is given by the following matrix
\begin{equation}
\label{c1}
C_1 = \left(
\begin{array}{ccc}
 e^{i (\gamma_2+\gamma_4)} c_{13} c_{23} & e^{i \gamma_2} c_{13} s_{23} & e^{-i \gamma_5} s_{13} \\
 -e^{i \gamma_4} c_{23} (1+s_{13}) s_{23} & c_{23}^2-s_{13} s_{23}^2 & e^{-i (\gamma_2+\gamma_5)} c_{13} s_{23} \\
 e^{i \gamma_5} \left(-c_{23}^2 s_{13}+s_{23}^2\right) & -e^{-i (\gamma_4-\gamma_5)} c_{23} (1+s_{13}) s_{23} & e^{-i (\gamma_2+\gamma_4)} c_{13} c_{23}
\end{array}
\right),
\end{equation}
which depends on five parameters $\gamma_2, \gamma_4, \gamma_5, \theta_{13}$ and $\theta_{23}$. The dispersion relations (\ref{omega}) now reads
\begin{equation}
\label{omega1}
\omega(k) = \arccos\left[ c_{13} c_{23}\cos(k-\gamma_2-\gamma_4)-\frac{1}{2} s_{23}^2(1+s_{13})\right].
\end{equation}

Notice that for
$$
\gamma_2=\gamma_4=\gamma_5 = 0,\quad \theta_{13} = \arcsin(1-\rho^2),\quad \theta_{23} = \arccos\left(-\frac{\rho}{\sqrt{2-\rho^2}}\right)
$$
the coin operator (\ref{c1}) reduces to
$$
C_\rho = \left(
            \begin{array}{ccc}
              -\rho^2 & \rho\sqrt{2-2\rho^2} & 1-\rho^2 \\
              \rho\sqrt{2-2\rho^2} & -1+2\rho^2 & \rho\sqrt{2-2\rho^2} \\
              1-\rho^2 & \rho\sqrt{2-2\rho^2} & -\rho^2 \\
            \end{array}
          \right).
$$
This is the family of coin operators we have found in \cite{iva:cont:def} through the deformation of eigenvectors of the Grover matrix.

\item $\delta \neq \gamma_1 - \gamma_2 - \gamma_4 $, $s_{13} = \frac{\sin(\gamma_1 - \gamma_2 - \gamma_4)}{\sin(\delta - \gamma_1 + \gamma_2 + \gamma_4)}$ - nontrivial solution

In this case, the coin operator equals
\begin{equation}
\label{c2}
C_2 = \left(
\begin{array}{ccc}
 e^{i \gamma_1} c_{23} B & e^{i \gamma_2} B s_{23} & -e^{-i (\delta+\gamma_5)} A \sin\kappa \\
 -e^{i (\gamma_1-\gamma_2)} A s_{23}c_{23} \sin\delta  & e^{i\kappa} \left(c_{23}^2+e^{i \delta} A s_{23}^2 \sin\kappa \right) & e^{-i (\gamma_1-\gamma_4+\gamma_5)} B s_{23} \\
 e^{i \gamma_5} \left(s_{23}^2+e^{i \delta} c_{23}^2 A \sin\kappa\right) & -e^{-i (\gamma_4-\gamma_5)} A s_{23}c_{23}\sin\delta  & e^{-i \gamma_1} c_{23} B
\end{array}
\right).
\end{equation}
For brevity we have used the notation
\begin{equation}
\label{kappa}
\kappa = \gamma_2+\gamma_4-\gamma_1,\quad A = \frac{1}{\sin(\delta+\kappa)},\quad B = \sqrt{A^2 \sin\delta \sin(\delta+2 \kappa)}.
\end{equation}
The set of solutions $C_2$ depends on six parameters, namely $\gamma_1, \gamma_2, \gamma_4, \gamma_5, \delta$ and $\theta_{23}$. We note that this class of coins is well defined only when the condition
\begin{equation}
\label{c2:cond}
-1\leq s_{13}\leq 1,\ {\rm i.e.}\ -1\leq\frac{\sin\kappa}{\sin(\delta + \kappa)}\leq 1,
\end{equation}
is satisfied. The dispersion relations are now determined by
\begin{equation}
\label{omega2}
\omega(k) = \arccos\left[B c_{23}\cos(k-\gamma_1) -\frac{1}{2}A s_{23}^2 \sin\delta \right].
\end{equation}

Notice that for the choice of the parameters
$$
\gamma_1 =\gamma_2 = \pi,\quad \theta_{23} = -\arctan 2,\quad \gamma_4=\gamma_5 = -\varphi,\quad \delta = \varphi + {\rm arccotan}\left(\frac{2\cot\varphi}{3}\right)
$$
the coin operator (\ref{c2}) reduces to
$$
C_\varphi = \left(
              \begin{array}{ccc}
                -\frac{\cos\varphi}{3} & \frac{2\cos\varphi}{3} & \frac{2\cos\varphi}{3}-i\sin\varphi \\
                \frac{2\cos\varphi}{3} & -\frac{\cos\varphi}{3}-i\sin\varphi & \frac{2\cos\varphi}{3} \\
                \frac{2\cos\varphi}{3}-i\sin\varphi & \frac{2\cos\varphi}{3} & -\frac{\cos\varphi}{3} \\
              \end{array}
            \right).
$$
This set is up to a global phase factor $e^{i\varphi}$ equal to the one-parameter family we have found in \cite{iva:cont:def}  by the deformation of eigenvalues of the Grover matrix.
\end{enumerate}

\end{enumerate}

The solutions $C_1$ and $C_2$ represents all coin operators which result in three-state quantum walk with point spectrum. Before we proceed with the analysis of their physical properties we point out that the derived results imply that the existence of point spectrum is a rather rare feature. Indeed, the found solutions depend on five, respectively six parameters, while a general coin operator depends on nine parameters. Hence, both families of coins $C_1$ and $C_2$ represent a set of zero measure in the unitary group $U(3)$.

\section{Peak velocities of the resulting quantum walks}
\label{sec4}

Let us now analyze the coin operators we have found in more detail. As a first physical parameter we consider the peak velocity \cite{kempf} which describes the rate of spreading of the quantum walk through the lattice. The peak velocity is determined as the maximum of the group velocity $v = \frac{d\omega}{dk}$. Notice that for both sets of coins $C_1$ and $C_2$ the dispersion relations are of the form
\begin{equation}
\omega(k) = \arccos\left(\rho\cos(k-\gamma) - \mu\right).
\end{equation}
To determine the peak velocity of the corresponding quantum walk we have to find $k_0$ such that the second derivative of $\omega$ vanishes. This leads us to the equation
\begin{equation}
\label{root}
\rho\mu(1+\cos^2(k_0-\gamma)) + (1-\rho^2-\mu^2)\cos(k_0-\gamma) = 0.
\end{equation}
The solutions are given by
\begin{equation}
\label{k0}
k_0 = \gamma \pm \arccos\Delta,
\end{equation}
where we have denoted
\begin{equation}
\Delta = \frac{\rho^2+\mu^2 - 1 + \sqrt{(1-\rho^2-\mu^2)^2-4\rho^2\mu^2}}{2\rho\mu}.
\end{equation}
The peak velocities are then found by evaluating the first derivative of $\omega$ at the point $k_0$. We obtain the following result
\begin{equation}
\label{v:peak}
v_{peak} = \frac{\rho\sqrt{1-\Delta^2}}{\sqrt{1-(\mu-\rho\Delta)^2}}.
\end{equation}
Note that neither $\Delta$ nor $v_{peak}$ depend on $\gamma$, so this parameter does not have a dynamical consequence.

\subsection{The class $C_1$}

For the set of coin operators $C_1$ the parameters $\rho$ and $\mu$ are given by
\begin{equation}
\rho = \cos\theta_{13}\cos\theta_{23},\quad \mu = \frac{1}{2}(1+\sin\theta_{13})\sin^2\theta_{23}.
\end{equation}
We find that for the five-parameter family of coins $C_1$ the peak velocity depends only on two, namely $\theta_{13}$ and $\theta_{23}$. The choice of the parameters $\gamma_2, \gamma_4, \gamma_5$ does not influence the dynamics of the quantum walk.

\begin{figure} [htbp]
\includegraphics[width=0.6\textwidth]{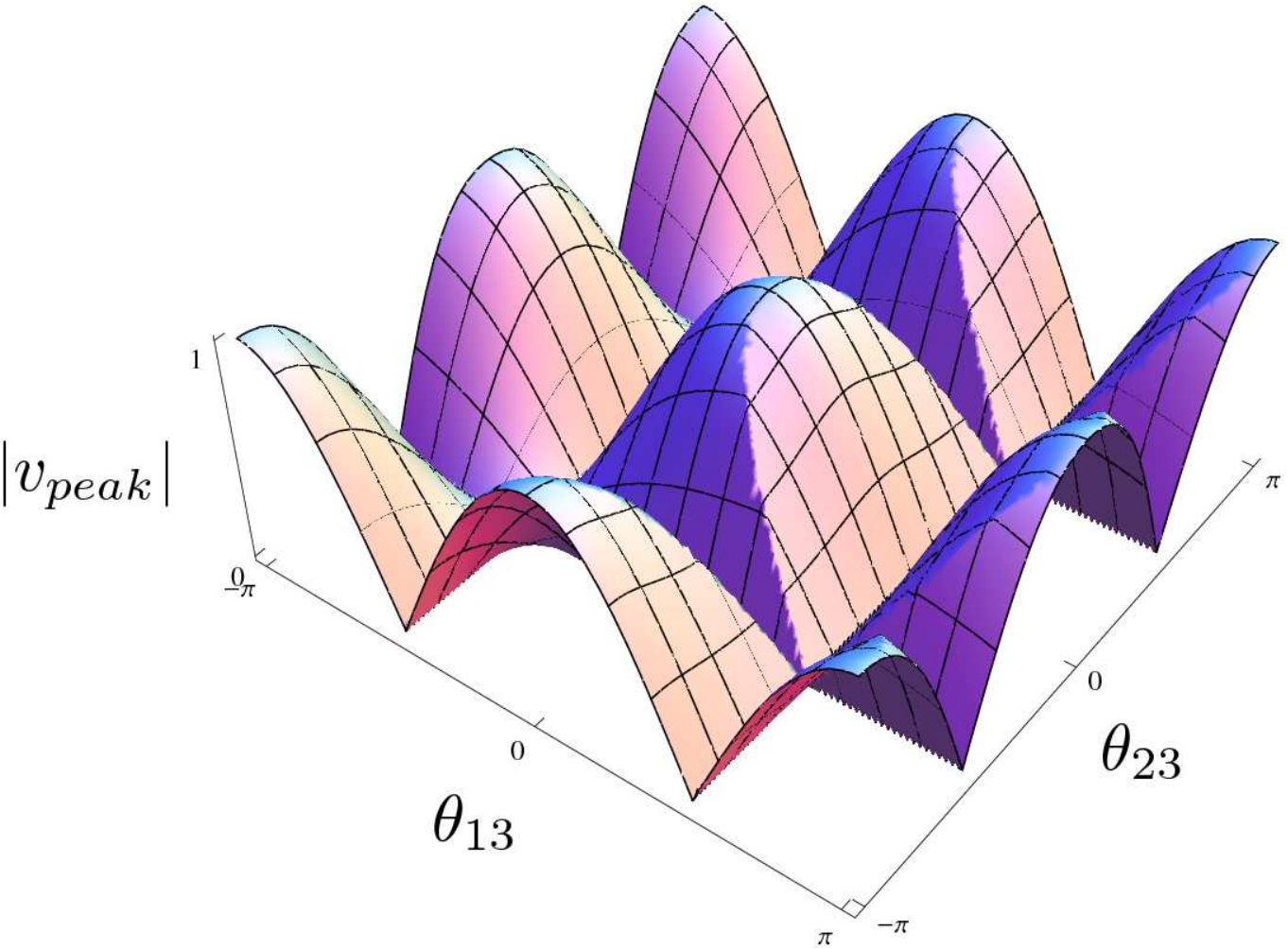}\hfill
\includegraphics[width=0.3\textwidth]{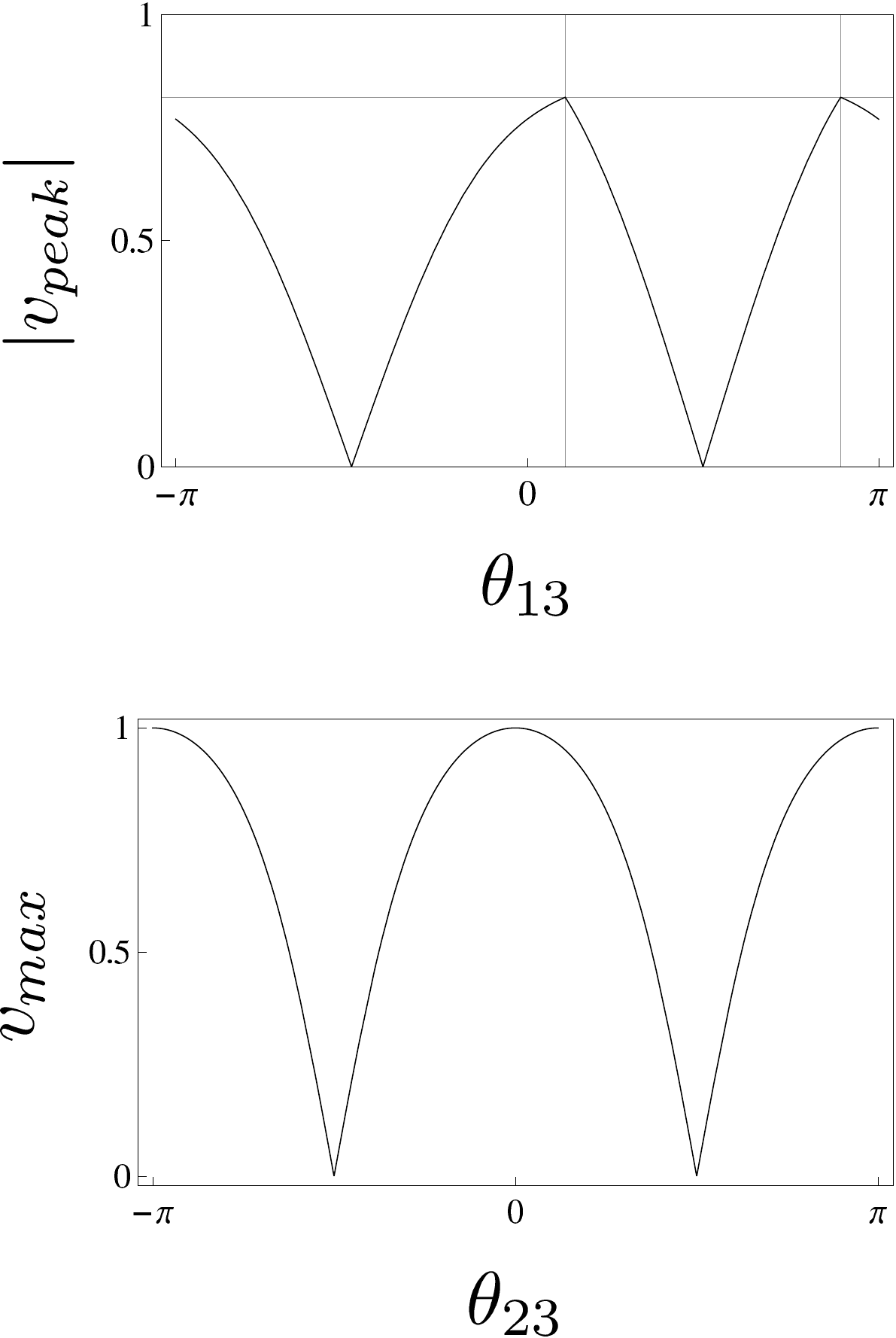}
\caption{On the left we show the absolute value of the peak velocity for the set of coin operators $C_1$ in dependence on the parameters $\theta_{13}$ and $\theta_{23}$. The upper right plot displays the cut for $\theta_{23}$ equal to $\pi/4$. Notice the two spikes at the values of $\theta_{13}$ determined by Eq. (\ref{curve}). The lower right plot shows the maximum peak velocity for a given $\theta_{23}$ which is determined by Eq. (\ref{vmax1}).}
\label{fig1}
\end{figure}

The peak velocity as a function of $\theta_{13}$ and $\theta_{23}$ is displayed in Figure~\ref{fig1}. As expected, the peak velocity is zero for $\theta_{13} = \pm\pi/2$ and $\theta_{23} = \pm\pi/2$, since these parameters correspond to the trivial solutions. The peak velocity is smooth except for the curve determined by
\begin{equation}
\label{curve}
\cos{\theta_{23}} = \pm\frac{\cos{\theta_{13}}}{1+\sin\theta_{13}},
\end{equation}
where it has a discontinuous derivative. For a given $\theta_{23}$ the maximum of the peak velocity lies on this curve and is given by
\begin{equation}
\label{vmax1}
v_{max} = \sqrt{\cos\theta_{23}\cos\left(2\arctan\left(\frac{1-\cos\theta_{23}}{1+\cos\theta_{23}}\right)\right)}
\end{equation}

\subsection{The class $C_2$}

In the second case the coefficients are equal to
\begin{equation}
\rho = \sqrt{\frac{\sin\delta \sin(\delta+2 \kappa)}{\sin^2(\delta+\kappa)} } \cos\theta_{23},\quad \mu = \frac{\sin\delta}{2\sin(\delta+\kappa)} \sin^2\theta_{23}.
\end{equation}
We see that for the six-parameter family of coins $C_2$ the peak velocity depends only on three, namely $\delta, \theta_{23}$ and $\kappa$ which is a linear combination of $\gamma_1, \gamma_2$ and $\gamma_4$, see equation (\ref{kappa}).

\begin{figure} [htbp]
\includegraphics[width=0.55\textwidth]{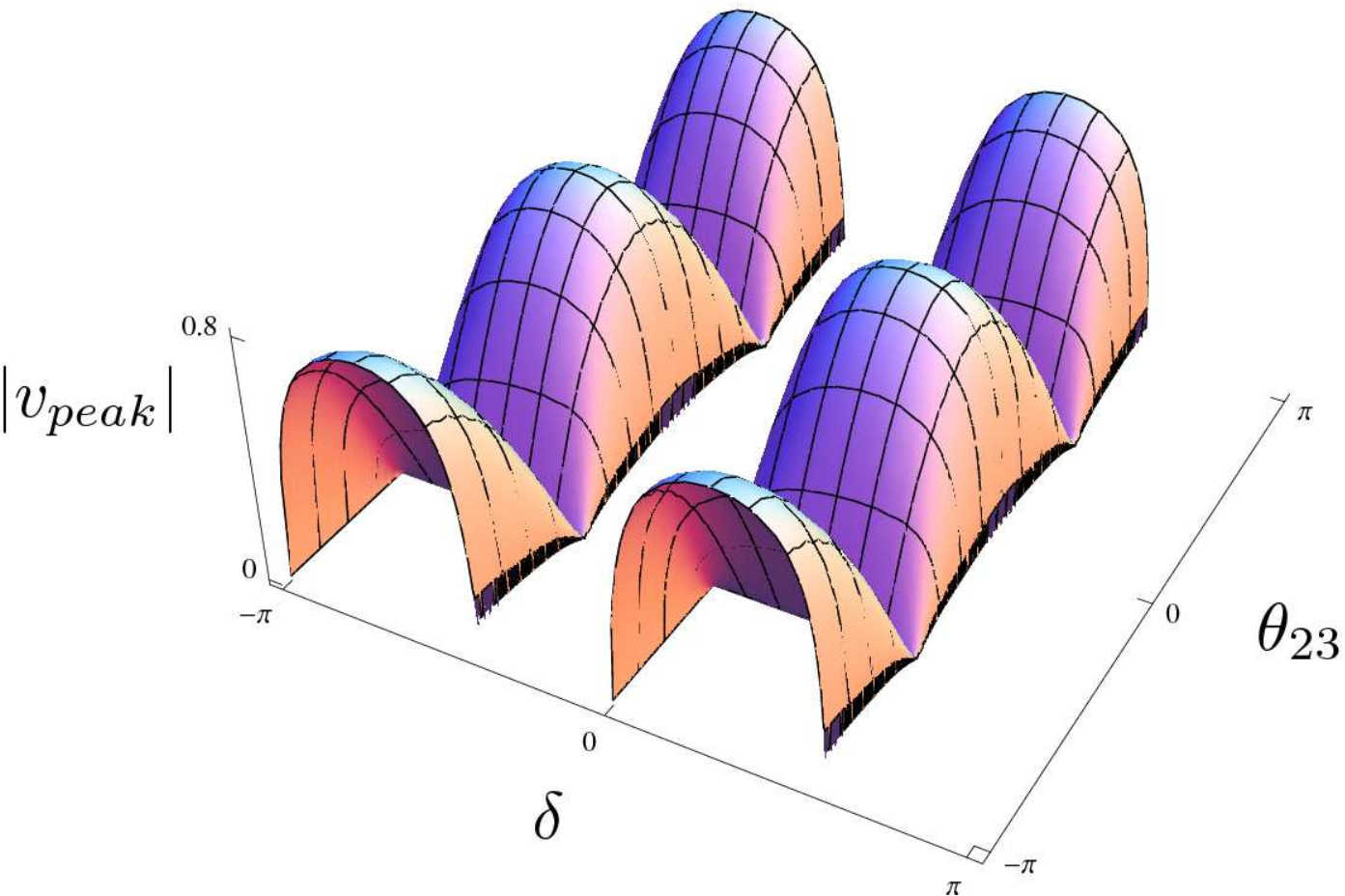}\hfill
\includegraphics[width=0.4\textwidth]{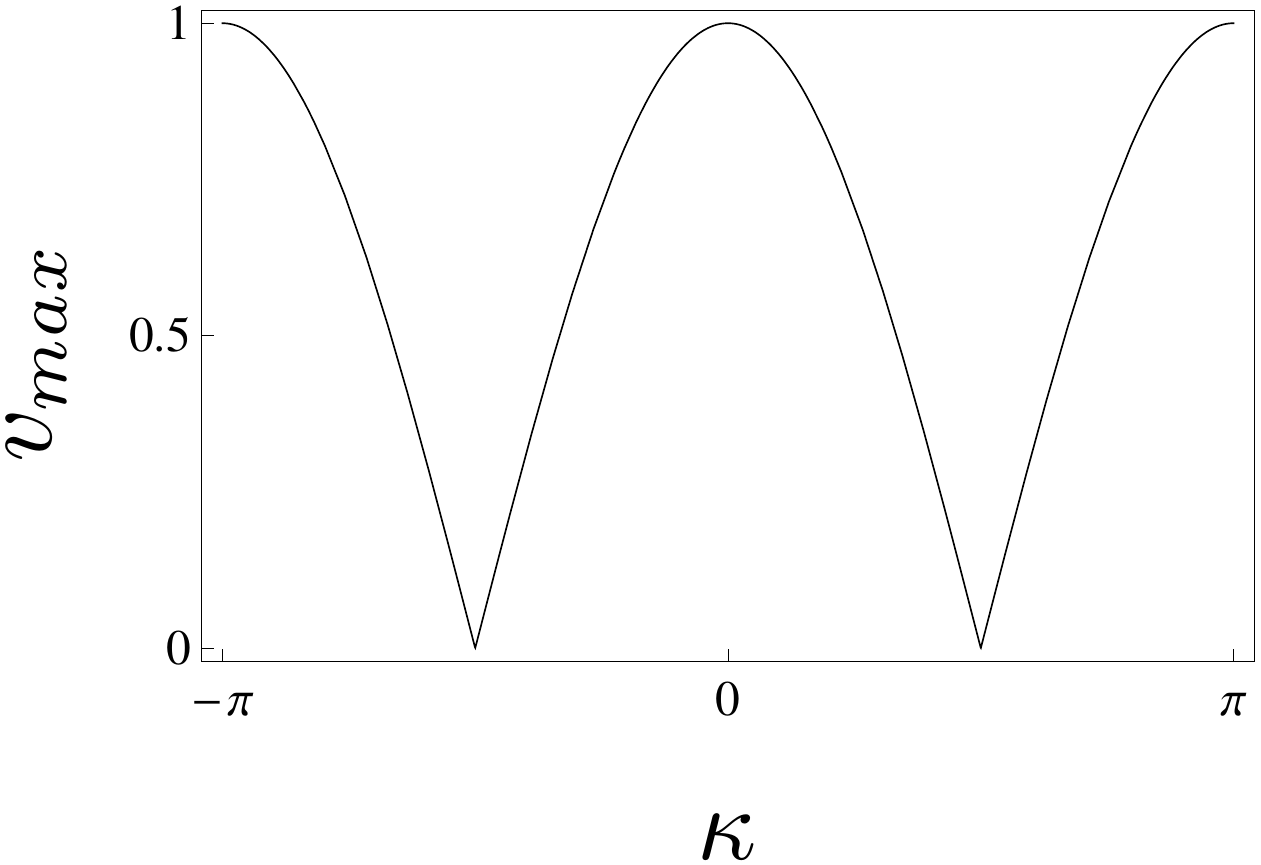}
\caption{On the left we show the peak velocity for the set of coin operators $C_2$. We have chosen the parameter $\kappa=\pi/5$. Due to the condition (\ref{c2:cond}) not all values of $\delta$ are admissible. The maximum of the peak velocity as a function of the angle $\kappa$ is shown on the right. The maximum is reached for the values $\theta_{23}=0$ and $\delta = \pi/2-\kappa$.}
\label{fig2}
\end{figure}

The peak velocity as a function of the angles $\delta$ and $\theta_{23}$ is shown in Figure~\ref{fig2} on the left. We fix the value of the remaining parameter $\kappa=\pi/5$. As before, the peak velocity vanishes for $\theta_{23} = \pm \pi/2$. On the right we display the maximum of the peak velocity as a function of $\kappa$. For a given $\kappa$, the maximum of the peak velocity is reached for $\theta_{23}=0$ and $\delta = \pi/2-\kappa$, and reduces to
\begin{equation}
v_{max} = |\cos\kappa|.
\end{equation}

\section{Trapping probability}
\label{sec5}

As we have already mentioned, the dynamics of the quantum walks we are interested in do not consist of just ballistic spreading. The existence of eigenvalue that is independent of $k$ and the corresponding bound state leads to partial trapping of the walker at the origin. In this section we analyze this feature in more detail.

We denote by $v(k) = \left(v^L(k), v^S(k), v^R(k)\right)^T$ the momentum representation of the (non-normalized) stationary state, i.e. the eigenstate of the evolution operator (\ref{evol:op}) corresponding to the constant eigenvalue. By $n(k)$ we denote the square norm of this vector. In a similar way, we denote by $v_{1,2}(k)$ the (normalized) eigenvectors of (\ref{evol:op}) corresponding to eigenvalues $e^{\pm i\omega(k)}$   Let the initial state of the coin be equal to
\begin{equation}
|\varphi\rangle = \varphi^L |L\rangle + \varphi^S|S\rangle + \varphi^R|R\rangle.
\end{equation}
The momentum representation of the initial state of the walk $|0\rangle\otimes|\varphi\rangle$ is then simply $\varphi=\left(\varphi^L, \varphi^S, \varphi^R\right)^T$. Using Fourier analysis \cite{ambainis} we find that the probability amplitude of particle being at the origin after $t$ steps of the walk is given by
\begin{eqnarray}
\label{psi:t}
\nonumber \psi(0,t) & = & \int\limits_{-\pi}^{\pi} \frac{dk}{2\pi}\ \frac{1}{n(k)} \left(v(k),\varphi\right)v(k) + \int\limits_{-\pi}^{\pi} \frac{dk}{2\pi}\ e^{i\omega(k) t} \left(v_1(k),\varphi\right)v_1(k) + \\
& &  + \int\limits_{-\pi}^{\pi} \frac{dk}{2\pi}\ e^{-i\omega(k) t } \left(v_2(k),\varphi\right)v_2(k).
\end{eqnarray}
With the stationary phase approximation \cite{statphase} one can show that the time-dependent integrals in (\ref{psi:t}) behave as $\sim t^{-\frac{1}{2}}$ for large values of $t$. Hence, in the limit $t\rightarrow + \infty$ only the first term in (\ref{psi:t}) remains and we find
\begin{equation}
\psi^\varphi_\infty \equiv \lim\limits_{t\rightarrow + \infty} \psi(0,t) =  \int\limits_{-\pi}^{\pi} \frac{dk}{2\pi}\frac{1}{n(k)} \left(v(k),\varphi\right)v(k).
\end{equation}
The localization probability is then equal to the square norm of the amplitude. Since we want to focus on the role of the coin operator on the walker trapping, we consider the initial coin state of the walker as the maximally mixed state. In such a case, the trapping probability can be expressed in the form
\begin{equation}
P_\infty = \frac{1}{3}\left(|\psi^L_\infty|^2 + |\psi^S_\infty|^2 + |\psi^R_\infty|^2\right),
\end{equation}
where $\psi^j_\infty$ is the limiting amplitude for the initial coin state $|j\rangle$, i.e.
\begin{equation}
\label{psi:j}
\psi^j_\infty = \int\limits_{-\pi}^{\pi} \frac{dk}{2\pi}\frac{1}{n(k)} {v^j}^*(k)v(k),\quad j=L,S,R.
\end{equation}
In the following we will see that $k$-dependence of the stationary state $v(k)$ involves only the term $e^{i k}$. The product ${v^j}^*(k)v(k)$ will be a linear combination of functions $e^{-i k}$, 1, $e^{i k}$. The square norm of the stationary state will be of the form
\begin{equation}
n(k) = a - 2b \cos(k-c).
\end{equation}
This implies that the amplitudes (\ref{psi:j}) can be decomposed into integrals
\begin{equation}
I_n = \int\limits_{-\pi}^{\pi} \frac{dk}{2\pi} \frac{e^{i n k}}{a-2b\cos(k-c)}, \quad n=-1,0,1.
\end{equation}
Such an integral can be turned into a contour integral over a unit circle in a complex plane
\begin{equation}
I_n = \left\{
        \begin{array}{c}
          e^{i k} = z \\
          dk = \frac{dz}{iz} \\
        \end{array}
      \right\} = -\frac{1}{2\pi i}\oint \frac{z^n dz}{be^{-ic} z^2 - az + be^{ic} },
\end{equation}
which is easily evaluated with the help of the residues. We find the following result
\begin{equation}
I_0 = \frac{1}{\sqrt{a^2-4b^2}},\qquad I_1 = I^*_{-1} = \frac{\frac{a}{\sqrt{a^2-4b^2}}-1}{2b} e^{i c}.
\end{equation}
Let us now specify the results for the two sets of coin operators.

\subsection{The Class $C_1$}

\begin{figure} [htbp]
\includegraphics[width=0.6\textwidth]{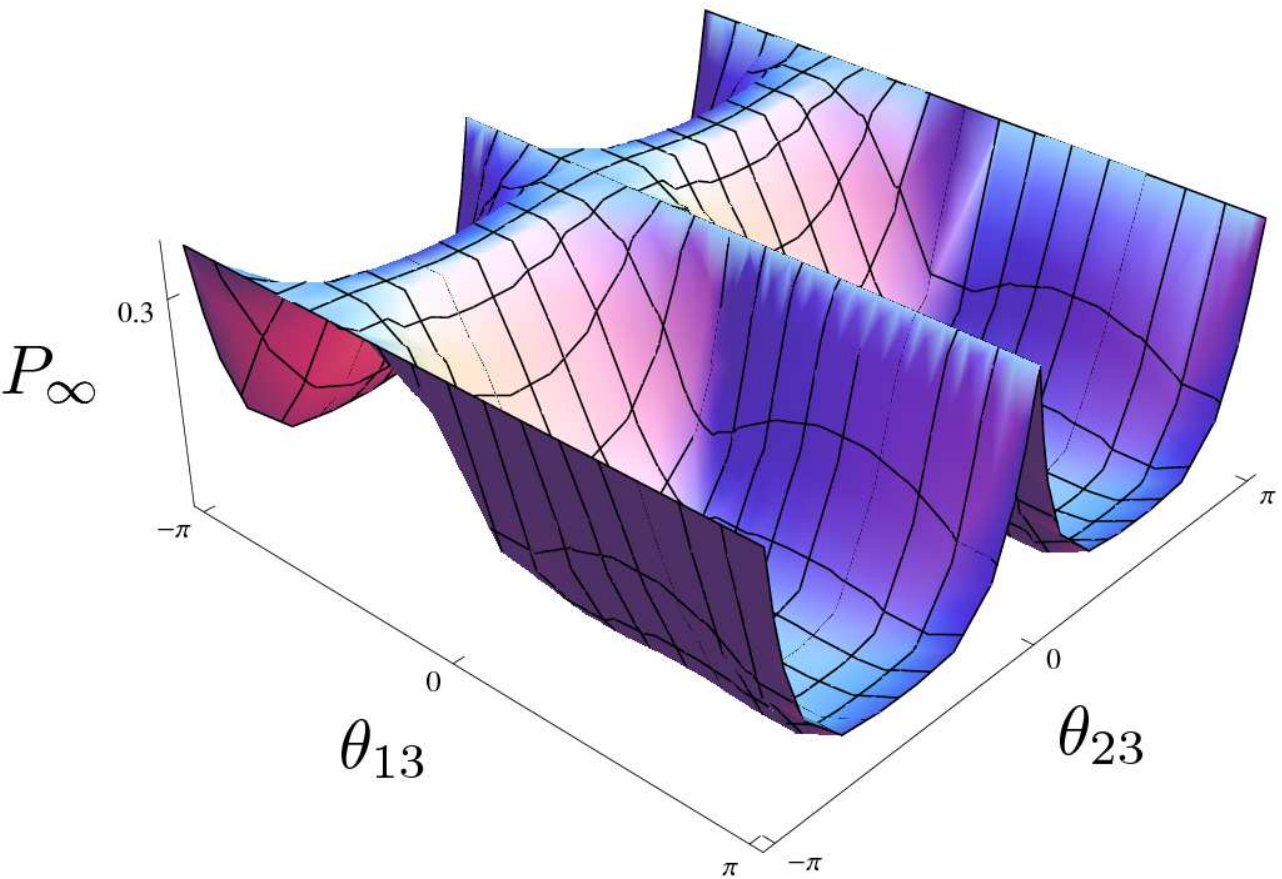}
\caption{Localization probability for the first class.}
\label{fig:3}
\end{figure}

For the first class of the coins the stationary state is given by
\begin{equation}
v(k) = \left(
       \begin{array}{c}
         -e^{-i\gamma_5}(\sin\frac{\theta_{13}}{2} + \cos\frac{\theta_{13}}{2})s_{23} \\
         e^{i(k-\gamma_2-\gamma_5)}(\sin\frac{\theta_{13}}{2} - \cos\frac{\theta_{13}}{2}) + e^{i(\gamma_4-\gamma_5)}(\sin\frac{\theta_{13}}{2} + \cos\frac{\theta_{13}}{2})c_{23} \\
          -e^{ik}(\sin\frac{\theta_{13}}{2} + \cos\frac{\theta_{13}}{2})s_{23} \\
       \end{array}
     \right).
\end{equation}
The square of the norm of this vector is equal to
\begin{equation}
n(k) = 2 + (1+s_{13})s^2_{23} - 2c_{13}c_{23}\cos(k-\gamma_2-\gamma_4).
\end{equation}
The parameters $a$, $b$ and $c$ are therefore
\begin{equation}
a = 2 + (1+s_{13})s^2_{23},\quad b = c_{13}c_{23},\quad c = \gamma_2+\gamma_4.
\end{equation}
We find that the limiting amplitudes at the origin (\ref{psi:j}) are given by
\begin{eqnarray}
\nonumber \psi^L_\infty & = & \left(
                            \begin{array}{c}
                              I_0 (1+s_{13})s^2_{23} \\
                              e^{-i\gamma_2}(|I_1| c_{13} - I_0c_{23}(1+s_{13}))s_{23} \\
                              e^{i\gamma_5} |I_1| (1+s_{13})s^2_{23} \\
                            \end{array}
                          \right), \\
\nonumber \psi^S_\infty & = & \left(
                            \begin{array}{c}
                             e^{i\gamma_2}(|I_1|c_{13}s_{23} - I_0)       \\
                             I_0 (1+c^2_{23} - s_{13}s^2_{23}) - 2 |I_1| c_{13}c_{23} \\
                             e^{i(\gamma_2 + \gamma_5)}(I_0 c_{13}s_{23} - |I_1| (1+s_{13})s_{23}c_{23})       \\
                            \end{array}
                          \right),\\
\psi^R_\infty & = & \left(
                            \begin{array}{c}
                            e^{-i\gamma_5}|I_1|(1+s_{13})s^2_{23} \\
                            e^{-i(\gamma_2+\gamma_5)}(I_0c_{13} - |I_1|(1+s_{13})s_{23}c_{23}) \\
                            I_0(1+s_{13})s^2_{23}\\
                            \end{array}
                          \right).
\end{eqnarray}
The trapping probability at the origin for a maximally mixed initial coin state then equals
\begin{eqnarray}
\nonumber P_\infty & = & \frac{1}{3} I_0^2 \left[1 + c^4_{23} + \frac{1}{2}(2+s^2_{23})^2 + (s^2_{13} + 2s_{13} - \frac{1}{2})s^4_{23}\right]-\\
\nonumber & & -\frac{4}{3} I_0 |I_1| c_{13}c_{23} \left[1 + c^2_{23} + (2 + s_{13}) s^2_{23}\right]-\\
 & & -\frac{4}{3} |I_1|^2 (1 + s_{13})(c^2_{23} s_{13}-1)
\end{eqnarray}
Notice that the result is independent of the $\gamma_i$'s. The only relevant parameters are the angles $\theta_{13}$ and $\theta_{23}$, i.e. the same parameters which also determine the rate of spreading of the walk. We display the behaviour of the localization probability in Figure~\ref{fig:3}.

\subsection{The Class $C_2$}

\begin{figure} [htbp]
\includegraphics[width=0.6\textwidth]{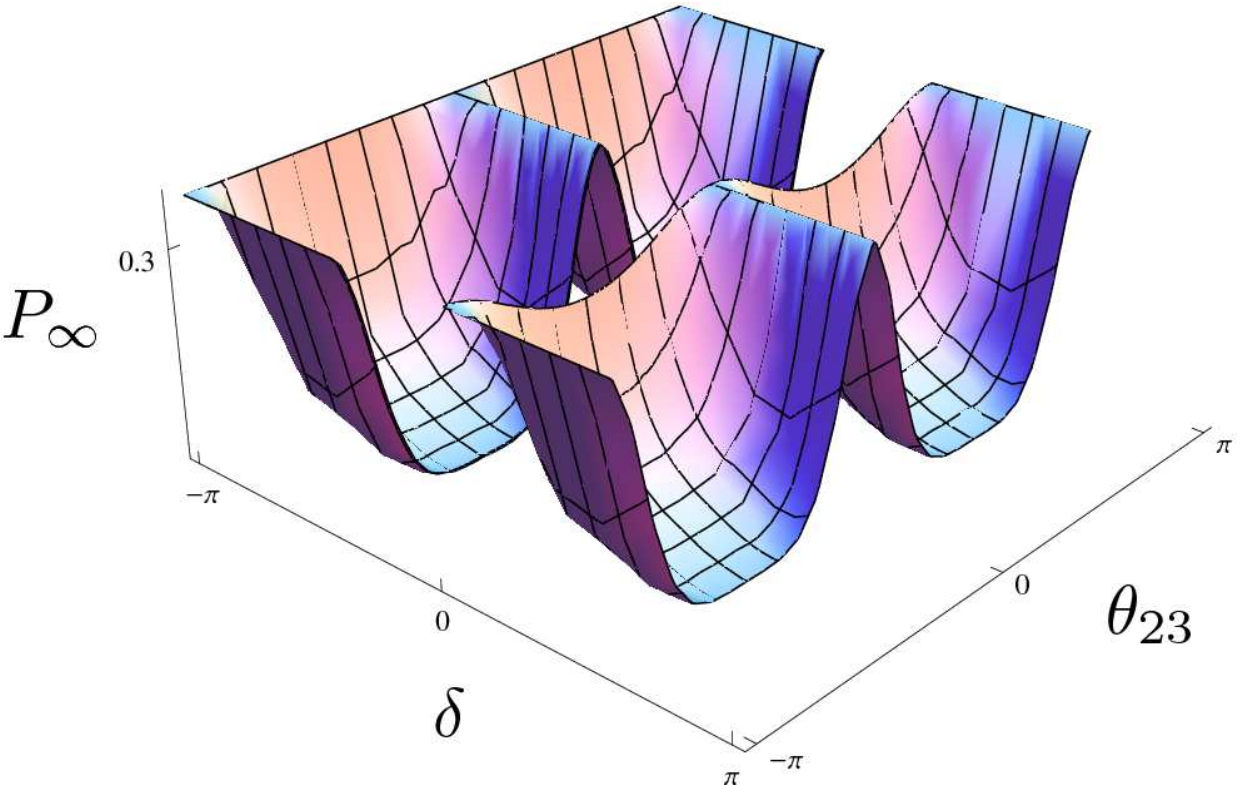}
\caption{Localization probability for the second class of coins. The parameter $\kappa$ has been chosen equal to $\pi/5$. The admissible range of the parameter $\delta$ is limited by the condition (\ref{c2:cond}).}
\label{fig:4}
\end{figure}

In the second case the stationary state is equal to
\begin{equation}
v(k) = \left(
        \begin{array}{c}
         e^{i(\gamma_2+\gamma_4)}\sin\delta s_{23}\\
         -e^{i(\gamma_1+\gamma_4)}\sin\delta c_{23} + e^{i(k+\gamma_4)}\sqrt{\sin\delta\sin(\delta+2\kappa)}   \\
         e^{i(k+\gamma_1+\gamma_5)}\sin\delta s_{13}\\
       \end{array}
     \right).
\end{equation}
The normalization of this vector is given by the factor
\begin{equation}
n(k) = \sin\delta\left(\sin(\delta+2\kappa)+(1+s^2_{23})\sin\delta - 2c_{23}\sqrt{\sin\delta\sin(\delta+2\kappa)}\cos(k-\gamma_1)\right)
\end{equation}
The parameters $a$, $b$, $c$ are then given by
\begin{equation}
a = \sin\delta (\sin\delta (1 + s^2_{23}) + \sin(\delta + 2 \kappa)),\quad b = \sin\delta c_{23} \sqrt{\sin\delta \sin(\delta + 2\kappa)},\quad c = \gamma_1.
\end{equation}
The probability amplitudes at the origin in the limit $t\rightarrow +\infty$ tend to the values
\begin{eqnarray}
\nonumber \psi^L_\infty & = & \left(
                                \begin{array}{c}
                                  I_0 \sin^2\delta s^2_{23} \\
                                  e^{i(\gamma_1-\gamma_2)}\sin\delta s_{23}(|I_1| \sqrt{\sin\delta \sin(\delta + 2\kappa)} - I_0c_{23}\sin\delta)\\
                                  e^{i(\gamma_1+\gamma_5-\kappa)} |I_1| \sin^2\delta s^2_{23} \\
                                \end{array}
                              \right), \\
\nonumber \psi^S_\infty & = & \left(
                                \begin{array}{c}
                                 e^{i(\gamma_2-\gamma_1)}\sin\delta s_{23}(|I_1| \sqrt{\sin\delta \sin(\delta + 2\kappa)} - I_0c_{23}\sin\delta) \\
                                 \sin\delta(I_0(c^2_{23}\sin\delta + \sin(\delta + 2\kappa)) - 2 |I_1| c_{23} \sqrt{\sin\delta \sin(\delta + 2\kappa)})\\
                                 e^{i (\gamma_1+\gamma_5-\gamma_4)}\sin\delta s_{23} (I_0 \sqrt{\sin\delta \sin(\delta + 2\kappa)} - |I_1| c_{23}\sin\delta)  \\
                                 \end{array}
                              \right), \\
\psi^R_\infty & = & \left(
                                \begin{array}{c}
                                e^{i (\kappa - \gamma_1-\gamma_5)} |I_1| \sin^2\delta s^2_{23} \\
                                e^{i (\gamma_4 - \gamma_1 - \gamma_5)} \sin\delta s_{23} ( I_0 \sqrt{\sin\delta \sin(\delta + 2\kappa)} - |I_1| c_{23} \sin\delta)  \\
                                I_0 \sin^2\delta s^2_{23} \\
\end{array}
                              \right).
\end{eqnarray}
Finally, the probability of finding the particle at the origin is given by
\begin{eqnarray}
\nonumber P_\infty & = & \frac{1}{3} I_0^2 \sin^2\delta\left[(c^2_{23}\sin\delta + \sin(\delta + 2\kappa))^2 + 2 \sin^2\delta s^4_{23} + \right.\\
\nonumber & &\left. + 2 \sin\delta s^2_{23}(c^2_{23} \sin\delta + \sin(\delta + 2\kappa)) \right] +\\
\nonumber & & + \frac{2}{3} |I_1|^2 \sin^3\delta\left[\sin\delta s^2_{23} + (1+c^2_{23})\sin(\delta+2\kappa)\right] - \\
& & -\frac{4}{3} I_0 |I_1| c_{23} \sin^2\delta \left[(1 + s^2_{23}) \sin\delta + \sin(\delta + 2\kappa) \sqrt{\sin\delta \sin(\delta + 2\kappa)}\right].
\end{eqnarray}
Note that the result depends on $\gamma_i$'s only through $\kappa = \gamma_2+\gamma_4-\gamma_1$. The localization probability thus depend only on $\delta$, $\theta_{23}$ and $\kappa$. These parameters also determine the peak velocity of the walk. We show the course of this function in Figure~\ref{fig:4}

\section{Conclusions}
\label{sec6}

We have found two classes of coins for three state quantum walks on the line which have a
point spectrum. Previously found coins \cite{iva:cont:def} having this property are special cases of the defined classes. In this perspective our results complete the classification of three-step quantum walks on a line which exhibit the localization effect. The obtained formulas (\ref{c1}), (\ref{c2}) determine all coin operators leading to localizing quantum walks. The sets of coin operators depend on five, respectively six parameters. Our results imply that localization is a rare feature, since both families of coins represent a set of zero measure in the unitary group $U(3)$.

Physical implications of our results have been discussed. As representative physical parameters we have chosen the propagation velocity and trapping probability at the origin.
We have shown that the peak velocity as well as the strength of localization depend only on few parameters defining the coin, namely two for the first class and three for the second. We derived explicit formula specifying the dependencies of the physically relevant parameters on the parameters used to define the coin matrix. Explicit formulas for the velocity and trapping allow to quantify the strength of localization and magnitude of the speed. Hence the extreme regimes of three state walks on the line can be pinpointed. The very moderate dependence of localization and peak velocity on matrix parameters has to be put into contrast with the original coin matrix which has nine independent parameters.

The identification of the two localizing coin classes allows us to estimate the degree of control we need to have over the coin in order to see localization in an experiment. Such an analysis is relevant when considering possible experimental implementations for instance using the optical feedback loop \cite{Schreiber}. The coin control, usually realized as an internal degree of freedom of the particle – spin or angular momentum \cite{and:2dwalk:science,craig}, must be sufficiently strict because localization exhibited by the coins applies only to a very limited range of parameters when compared to the full parameter size of a general $U(3)$ coin.

\begin{acknowledgments}
The financial support from RVO 68407700, SGS13/217/OHK4/3T/14, GA\v CR 13-33906S and GA\v CR 14-02901P is gratefully acknowledged.
\end{acknowledgments}

\end{document}